\documentclass[11pt]{article}
\usepackage[noadjust]{cite}
\usepackage[top=2cm, bottom=2cm, left=2cm, right=2cm]{geometry}

\usepackage{hyperref}
\usepackage{amsmath,amssymb,amsfonts}
\usepackage{algorithmic}
\usepackage{graphicx}
\usepackage{textcomp}
\usepackage[figuresright]{rotating} 
\usepackage{color, colortbl}
\usepackage{gensymb}
\usepackage{float}
\usepackage{placeins}
\definecolor{Gray}{gray}{0.9}
\definecolor{LightCyan}{rgb}{0.88,1,1}

\def\BibTeX{{\rm B\kern-.05em{\sc i\kern-.025em b}\kern-.08em
    T\kern-.1667em\lower.7ex\hbox{E}\kern-.125emX}}
\begin{document}

\title{Lightning Mapping: Techniques, Challenges, and Opportunities}

\author{Ammar Alammari\footnote{Institute of Sustainable Energy, Universiti Tenaga National, Jalan Ikram-Uniten, 43000 Kajang, Selangor, Malaysia}\textsuperscript{$\hspace{5pt}\dagger$},
 Ammar Ahmed Alkahtani\textsuperscript{*}, 
 Mohd Riduan Ahmad\footnote{Atmospheric and Lightning Research Laboratory, Center for Telecommunication Research and Innovation (CeTRI), Fakuliti Kejuruteraan Elektronik dan Kejuruteraan Komputer (FKEKK), Universiti Teknikal Malaysia Melaka (UTeM), Melaka, 76100, Malaysia},
 Fuad M. Noman\textsuperscript{*},\\
 Mona Riza Mohd Esa\footnote{IVAT, Sekolah Kejuruteraan Elektrik, Fakulti Kejuruteraan, Universiti Teknologi Malaysia (UTM), Johor, 8300, Malaysis}, 
 Zen Kawasaki\footnote{Graduate School of Engineering, Osaka University, 1-1 Yamadaoka, Suita, Osaka 565-0871, Japan},
 Sieh Kiong Tiong\textsuperscript{*}
}

\markboth{To be Submitted for review}%
{Shell \MakeLowercase{\textit{et al.}}: Bare Demo of IEEEtran.cls for Journals}

\maketitle

\begin{abstract}
Despite the significant progress made in studying the lightning phenomenon, precise location and mapping of its occurrence remain a challenge. Lightning mapping can be determined by studying the electromagnetic radiation accompanying the lightning discharges. It can contribute substantially to efforts made to protect lives and valuable assets. There are three main methods used to locate lightning, which are Magnetic Direction Finder (MDF), Time of Arrival (TOA), and Interferometer (ITF). A thorough study of these methods provides researchers with a guide to better understand and progress in this field.  This paper reviews existing approaches used to locate and map lightning within these three methods. We study the implemented techniques, analyze their merits and demerits, and sort them in a way that facilitates extracting opportunities for further improvements. We conclude that for better development in determining the location and map of lightning, improving the processing of lightning signals and filtering the associated noise with it is essential. This includes introducing new processing methods such as wavelet transformation instead of the traditional cross-correlation. The use of artificial intelligence may also contribute a lot, particularly deep learning, to determining the type of lightning, which enables better mapping for the lightning and its occurrence. We also could conclude that unlike MDF and TOA, which can locate the lightning strike points, ITF can produce lightning discharge propagation images that can unveil the mechanism of lightning discharges. Finally, this paper serves as a reference for researchers focusing on lightning mapping to give them insight into the field.
\end{abstract}

\vspace{-0.02in}

\vspace{-0.05in}

\section{Introduction} 
\label{sec:introduction}
Lightning is a natural phenomenon of electrical discharges that occurs between two different polarity objects of cloud and ground, two clouds, or within a cloud. When discharges are generated, electromagnetic radiations over different ranges of frequencies are produced, usually extending from the ultra-low frequency (ULF) through the ultra-high frequency (UHF) \cite{Cummins1998}. Lightning discharges are mainly categorized into two types, cloud-to-ground (referred to as CG, including downward negative, upward negative, downward positive, and upward positive) and cloud discharges (usually referred to as IC, including intracloud, intercloud, and cloud-to-air) \cite{Rison1999, Shao1996, Sun2013, Thomas2001, Zhang2012}. Although the physics behind lightning initiation remains unclear, many hypotheses have been proposed in the literature. Two candidate theories are considered during the lightning initiation process which are hydrometeor-initiated positive streamers and cosmic ray-initiated runaway breakdown \cite{Petersen2008}. When the electric field between charges becomes sufficiently large, the lightning is initiated. The massive amount of electromagnetism generated makes lightning a major cause of electromagnetic interference that can affect various electronic systems. Lightning also is one of the major causes of death in various countries around the world. To improve the protection for both human and valuable assets, different lightning mapping systems have been introduced long ago. However, one major challenge of these old systems is the processing time where most of the work is done on an offline basis \cite{Akita2014, Rison1999, Stock2014, Ushio2011, Zeng2016}. With the technological advancements made in signal processing, these systems could now be able to make significant progress in real-time detection of lightning strikes and lead to mitigating the impact of lightning.

Lightning mapping systems are generally categorized into two methods, the three-dimensional (3D) (e.g. lightning mapping arrays (LMA)) and the two-dimensional (2D)  mapping (e.g. interferometer--ITF) \cite{Abbasi2019}. The LMA is usually composed of 6 to 20 VHF antennas separated by a distance of kilometres and operates at different frequency ranges. The ITF systems, on the other hand, are composed of 3-4 VHF antennas within several meters distance and operate at selected frequency ranges \cite{Abbasi2019, Stock2014}. The ITF is superior to LMA in which it is capable of scanning more lightning events than that of LMA and it performs a continuous and quasi-continuous emission mapping from lightning pulses for better 2D visualization \cite{Rison1999}. Modern lightning mapping systems have been designed to operate in different frequency bands ranging from extremely low frequency (ELF) to VHF frequencies. VHF lightning mapping was traditionally performed using the time of arrival (TOA) technique \cite{Tantisattayakul2005}, developed in Florida and resulting in new 3D accurate measurements to locate the RF radiation sources \cite{Lennon1991, Proctor1971, Proctor1984}. The method uses the TDOA technique that requires a minimum of four to five antennas. On the other hand, the use of ITF in determining the DOA has been developed and enhanced over the past forty years by many researchers \cite{Rhodes1989, Hayenga1981, Qiu2009, Rhodes1994, Warwick1979}.  This evolved into the current lightning mapping using ITF methods that allow for a more accurate mapping of the lightning. The ITF can locate a source of VHF impulse based on the digital interferometric technique. In the latest development of ITF \cite{Stock2014} the system receives VHF signals from different antennas that work as an array within a specific distance of a few wavelengths (10-20 m and more). The system can capture the electric field changes caused by the lightning discharge in the VHF band. The system consists of three resistively coupled of flat plate antennas arranged to form two equally spaced length orthogonal baselines along with the fourth antenna to record the time series data. This system enables the 3D mapping of sources in the azimuth and elevation angles. 

On the other hand, VLF has been used in a wide range of research areas including magnetic direction finder (MDF) and TOA. In MDF, the objective is to find the location from which the lightning initiated. However, the accuracy of the conventional VLF-MDF is relatively poor in detecting less distant lightning (below 200 km), which depends on the spacing of the sensors \cite{Cummins2000}. Several attempts of using the gated technique have been proposed to improve the accuracy of narrowband VLF-MDF at short ranges. Krider \textit{et al.} \cite{Krider1976} introduced the first commercial gated wideband MDF system to overcome the problem of large errors in narrowband MDF. The gated wideband DF provides improved azimuthal errors achieving a mean value of 1$\degree$ and a standard deviation of 2$\degree$ \cite{Krider1976}. 

It is worth mentioning that the key factors in differentiating between mapping techniques include the number of sensors (i.e. antennas) required to find the lightning location, baselines between sensors, network geometry, and the type of sensors deployed \cite{Dong2002, Shao1995, Zhang2008}. In some of the systems, a combination of two mapping techniques is used to improve the accuracy as well as to cover wider mapping distance in kilometres with minimizing the number of sensors (e.g. antennas) baselines \cite{Cummins2000}. Kawasaki \cite{Kawasaki2012}, attempted to give a clear image and summary of both systems of the TOA lightning techniques compared to the ITF and how they are equivalent to each other. The study distinguished both systems from different perspectives based on the pulse radiation procedures linked to the lightning discharge and its location by discussing the principles of each system accurately. However, the combination of TOA/ITF is still in the developing stage to operate in real-time.

\section{SOURCES AND METHODS}
\begin{figure}[!h]
	\centering
	\includegraphics[width=0.7\linewidth,keepaspectratio]{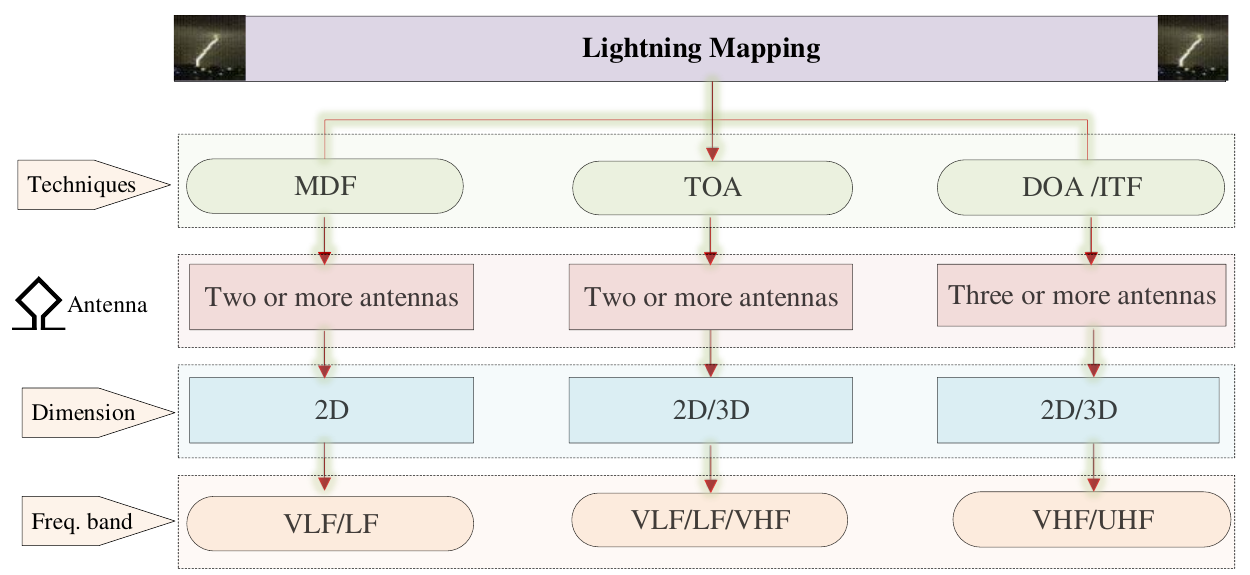}
\caption{Lightning Mapping Methods.}
\label{Fig:Fig1}
\end{figure}

Figure~\ref{Fig:Fig1} shows a summary of the commonly used lightning mapping methods which are being reviewed in this paper. The methods are further categorized according to the number of antennas, type of baselines, number of dimensions to which the lightning is mapped, and frequency bands. This classification enables the reader to understand the taxonomy used throughout this paper. 

\begin{figure}[!h]
	\centering
	\includegraphics[width=0.7\linewidth,keepaspectratio]{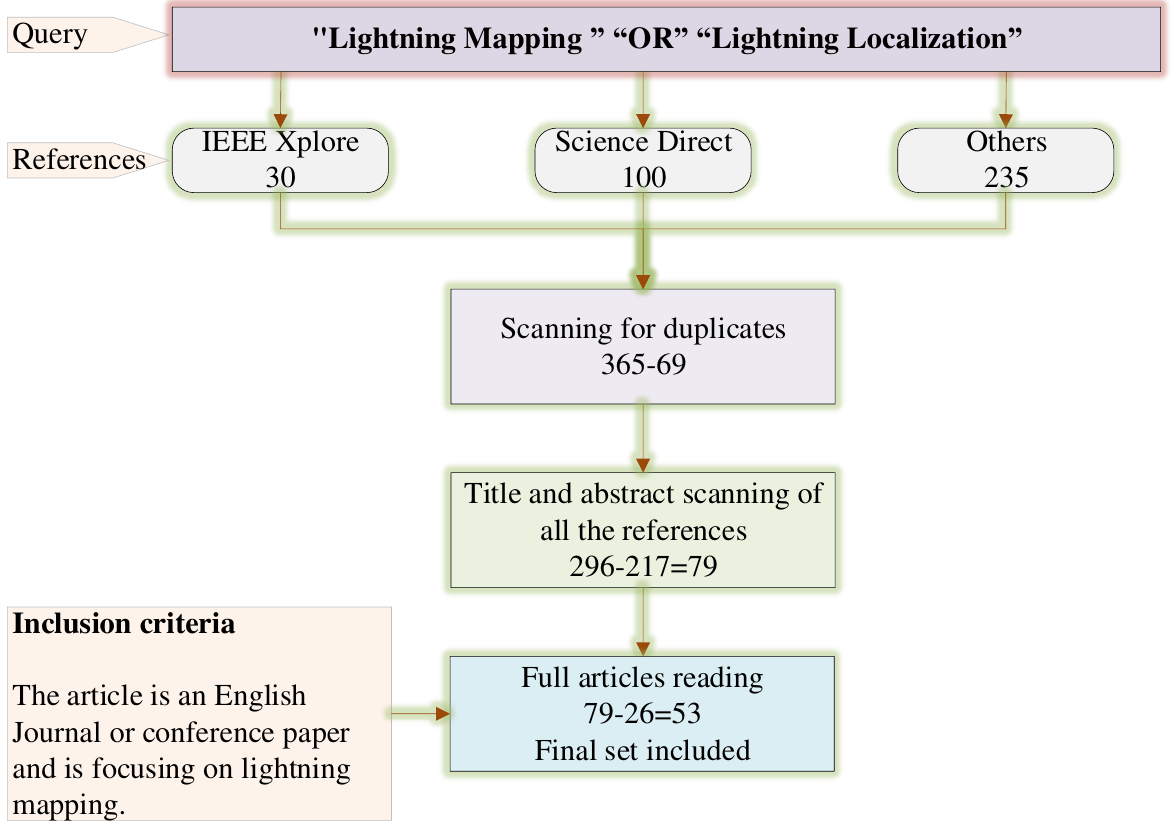}
\caption{Paper query scheme.}
\label{Fig:Fig2}
\end{figure}

From the queries in Figure~\ref{Fig:Fig2}, we limit our search query to lightning mapping, and lightning localization because these are the most important keywords in this scope of research. To further, simplify the reviewing process, a full list of the resulting papers from the search, together with their corresponding categories based on the taxonomy in Figure~\ref{Fig:Fig2}, which are compiled from various sources in the literature into a Microsoft Word format. In this respect, comprehensive reading is performed, resulting in a large collection of highlights and comments regarding related papers on this topic. We classify studies based on the measurement setups and techniques used for filtering and processing the lightning data, as well as on the mapping techniques implemented. To the best of our knowledge, no review has been done on these parameters; thus, this paper will serve as a guideline for researchers focusing on mapping lightning using MDF, TOA, and DOA techniques.  

The articles were classified based on the initial taxonomy, where all comments on the body of texts were saved according to the preference of the authors (either hardcopy or softcopy version). This step eases the process of summarization, tabulation, and a description of the main findings. Sets of relevant information were also saved in Microsoft Word and Microsoft Excel formats, including a list of full articles, together with the respective source database. A summary and description tables and categorization tables based on lightning localization and lightning mapping were presented in this study.

Based on Figure~\ref{Fig:Fig2}, a primary query was conducted in Google Scholar and resulted in 365 different articles; 30 were collected from the IEEE Explore database, 100 from the ScienceDirect database, and 235 from other sources. Initially, we checked all downloaded articles for possible duplicates, from which 69 articles were removed. During the scanning process of titles and abstracts, 217 additional articles were excluded, as these were not related to the main topic of study. We then conducted full-text reading on the remaining articles from which 26 articles were classified as out of the scope of lightning mapping. The exclusion of these articles leaves only 53 articles in the final set for the review. Hence, these papers were analyzed thoroughly to determine a general aim to direct the research on this topic. The selected articles are strictly related to the actual lightning mapping literature or studies describing existing works on lightning mapping.

\section{MAGNETIC DIRECTION FINDER}
The basic principle of the MDF is that it includes the use of two vertical, orthogonal loops with planes oriented at NS, and EW, each measuring the magnetic field from a given vertical radiator, can be used to obtain the direction to the source. However, some of the disadvantages of using MDF are the use of single station-(one loop antenna) which can only determine the azimuth angle of the lightning source \cite{Murty1973}. It was realized that the MDF method requires at least two stations to determine lightning location sources. The azimuth angle is calculated between the station, the South-North (SN) plane, and the direction where the lightning strikes. A line is drawn from the station to cross the unknown lightning strike point. At least two lines are then drawn from two stations, with the intersection between these two lines representing the exact lightning strike location, as shown in Figure 3. Therefore, to reduce the estimation error of the lightning strike location, several stations are required. In the MDF method, some cases could involve a lightning strike hitting a point on the line between the stations, making it very difficult to determine the location of the lightning strike. In this situation, at least three stations are required to observe the lightning strike location.

\begin{figure}[!h]
	\centering
	\includegraphics[width=0.7\linewidth,keepaspectratio]{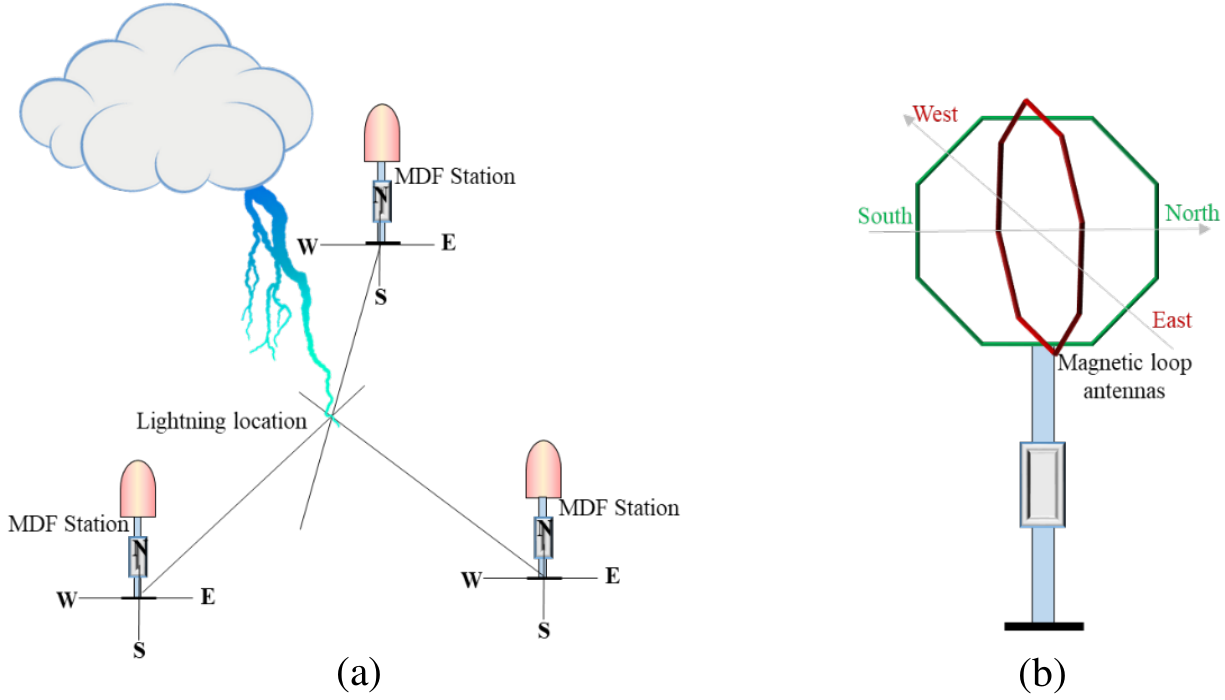}
\caption{Lightning detection using MDF method. (a) multi-station MDF; (b) Orthogonal magnetic antenna.}
\label{Fig:Fig3}
\end{figure}

MDF is a method that uses the magnetic field to locate, track, and measure the intensity of lightning emissions. It is mainly used to find the azimuth angle of return stroke in CG lightning. The MDF systems are implemented by using a vertical magnetic loop or cross-loop antenna \cite{Lojou2009, Lu2020}. The MDF technique requires at least two sensors to determine the lightning strike location. The sensors can detect the azimuth angle, which is located between the north direction and the direction of the lightning strike \cite{Cummins2000}.

MDF has been applied for estimating the peak the lightning strikes from the main radiating sources to pinpoint retriggering flashes. However, due to the errors in previous measurements for estimating return strike points in early versions of MDF, other researchers \cite{Uman1980} proposed a theory to minimize the measurement errors by implementing two baselines with a sampling frequency of (1 kHz to 1 MHz) based on the VLF bands. The study resulted in a system with poor accuracy due to the close ranges between the VLF measurements. Another study by Orville \cite{Orville1991} attempted to optimize the accuracy of MDF of CG lightning flashes operate with a range of frequency (10 kHz to 3MHz). The statistical analysis involved was based on the geometrical of magnetic finding. The system resulted in a detection error reduction by 80\% from those in previous measurements \cite{Orville1991, Sonnadara2000, Tao2018}. Sonnadara \textit{et al.} \cite{Sonnadara2000} studied the reconstruction of the lightning strike locations based on CG lightning flashes by using two wideband MDF stations. The optimization of the direction finder instrument uncertainty localization accuracy of $\pm 5$ km within a 100 km radius was produced. 

Orville \cite{Orville1991} used MDF to estimate the peak current of return strokes triggered of CG flashes in Georgia and Florida by estimating and calculating the exact radiated lightning. The system was able to observe six MDFs of each an orthogonal magnetic loop and a flat plate antenna was used with a bandwidth range of (1-350 kHz). The study continuously measured the lightning locations and protections of 18-triggered lightning return strokes, which were collected in four years. The peak current of lightning stroke was estimated by using normalized signal strength amplitudes strokes. The sensitivity of the evaluation-produced errors of lightning peak errors was 2\%. The study also suggests that the MDF technique could become a suitable method for enhancing the estimation of lightning return strokes in 2D with improved estimation of distance, rise time, and peak current. 

Yang \textit{et al.} \cite{Yang2010} suggested some improvements to the MDF technique in measuring the closed magnetic field in the triggering flashes in 2D. The study focused on the statistical distribution of the used channel and base currents to better analyze the closed magnetic field. The authors used numerical methods to examine the effects of different parameters, such as current rise time, return stroke speed, distance, and peak current on the closed magnetic field. They concluded that the radiation components of the total magnetic field peaks depend on the assumption of the current rise time and the return stroke speed. Generally, the magnetic field peak does not depend on the peak current but strongly depends on the distance from the radiation source. The distance effects on the time variation in the radiation components are quite different from that of the return stroke speed and the current rise time. In this case, increased distance and rise time causes a decrease in the field peak value, while there is an increase in the rise time.

Tao and Lihua \cite{Tao2018} investigated two MDF stations operating based on two baselines to reconstruct the position the CG lightning flashes. They performed numerical simulations to find the relation between errors of the calculated lightning striking points and the location of the stations. It was found that the locations of the stations are heavily dependent on the orientation of the striking points.
 
Table~\ref{tab1} summarizes the MDF related methods that have been reviewed. The table shows details of the configuration and parameters of the design of MDF systems. The table also presents the advantage and disadvantages of each study.

\hphantom{
\cite{Uman1980}
\cite{Orville1991}
\cite{Sonnadara2000}
\cite{Cummins1998}
\cite{Tao2018}
}
\vspace{-0.18in}
\begin{table*}
\caption{Summary of MDF methods from the literature.}
\begin{tabular}{c}
	\begin{minipage}{1\textwidth}
      \includegraphics[width=\linewidth, height=7.5in]{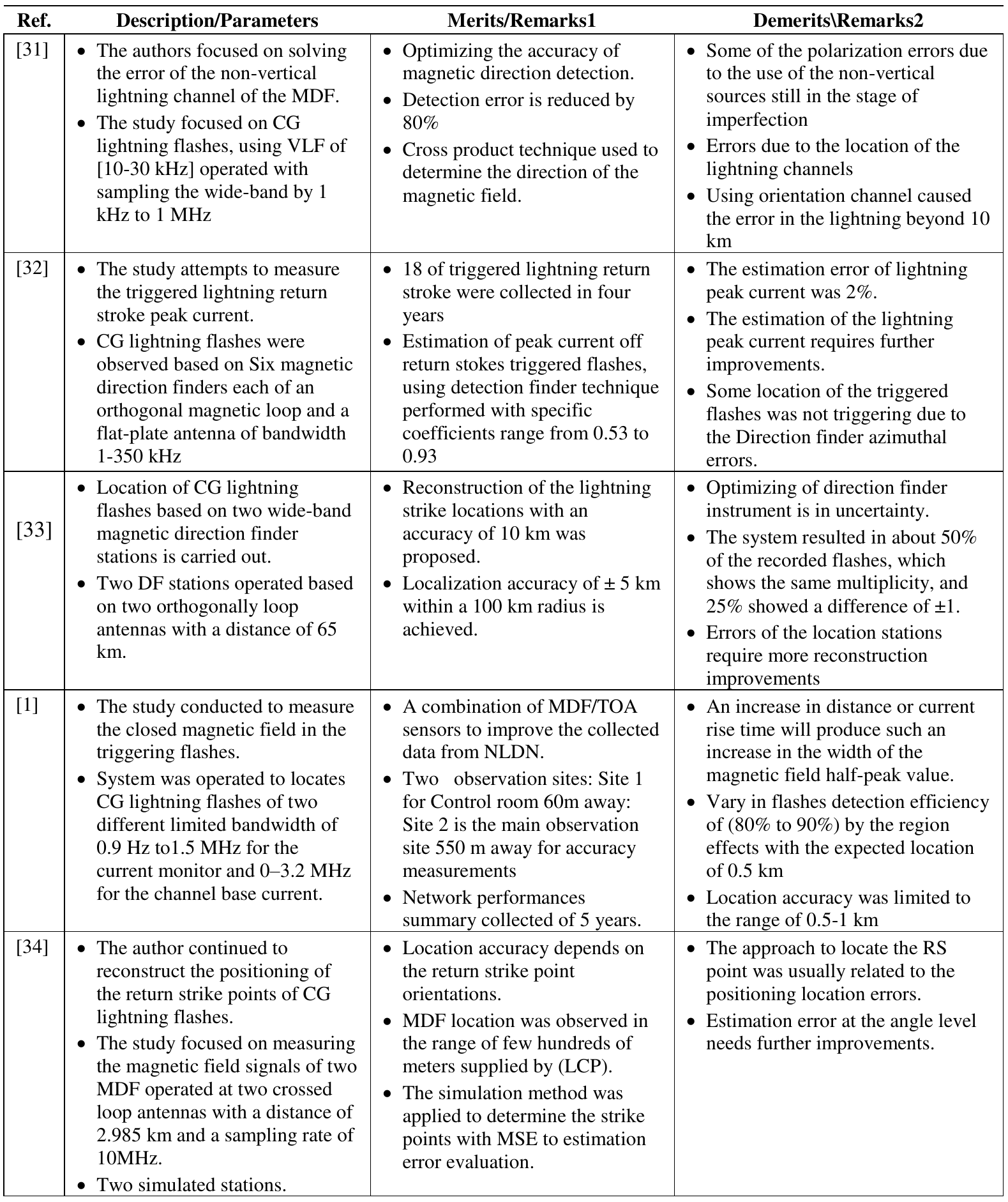}
    \end{minipage}
\end{tabular}
\label{tab1}
\end{table*}

\section{TIME OF ARRIVAL (TOA)}
Generally, lightning mapping systems use the TOA technique to map lightning strikes through the recorded VHF pulses associated with lightning discharges. These systems can be divided into three levels: (1) the very short-baseline technique operating at a distance of tens to hundreds of meters, (2) the short-baseline technique operating at a distance of hundreds to thousands of kilometres (km) \cite{Cianos1972, Taylor1978} and (3) the long-baseline technique operating at a distance of hundreds to thousands of km with a VLF/LF band \cite{Lewis1960}.

The basic principles of the TOA method is that it require at least three widely separated stations (few to tens of kilometres) to determine the lightning location. Adding more than three stations will give a more precise analysis of the measured arrival times of the TOA impulsive VHF events. Figure 4 illustrates a simplified setup of the TOA locating system. In this example, four lightning stations are used (1 to 4), where each station independently records the lightning events and use a peak detector to determine the instantaneous time of these events. This could lead to different recorded times at each station, which degrades the localization performance.

\begin{figure}[!h]
	\centering
	\includegraphics[width=0.7\linewidth,keepaspectratio]{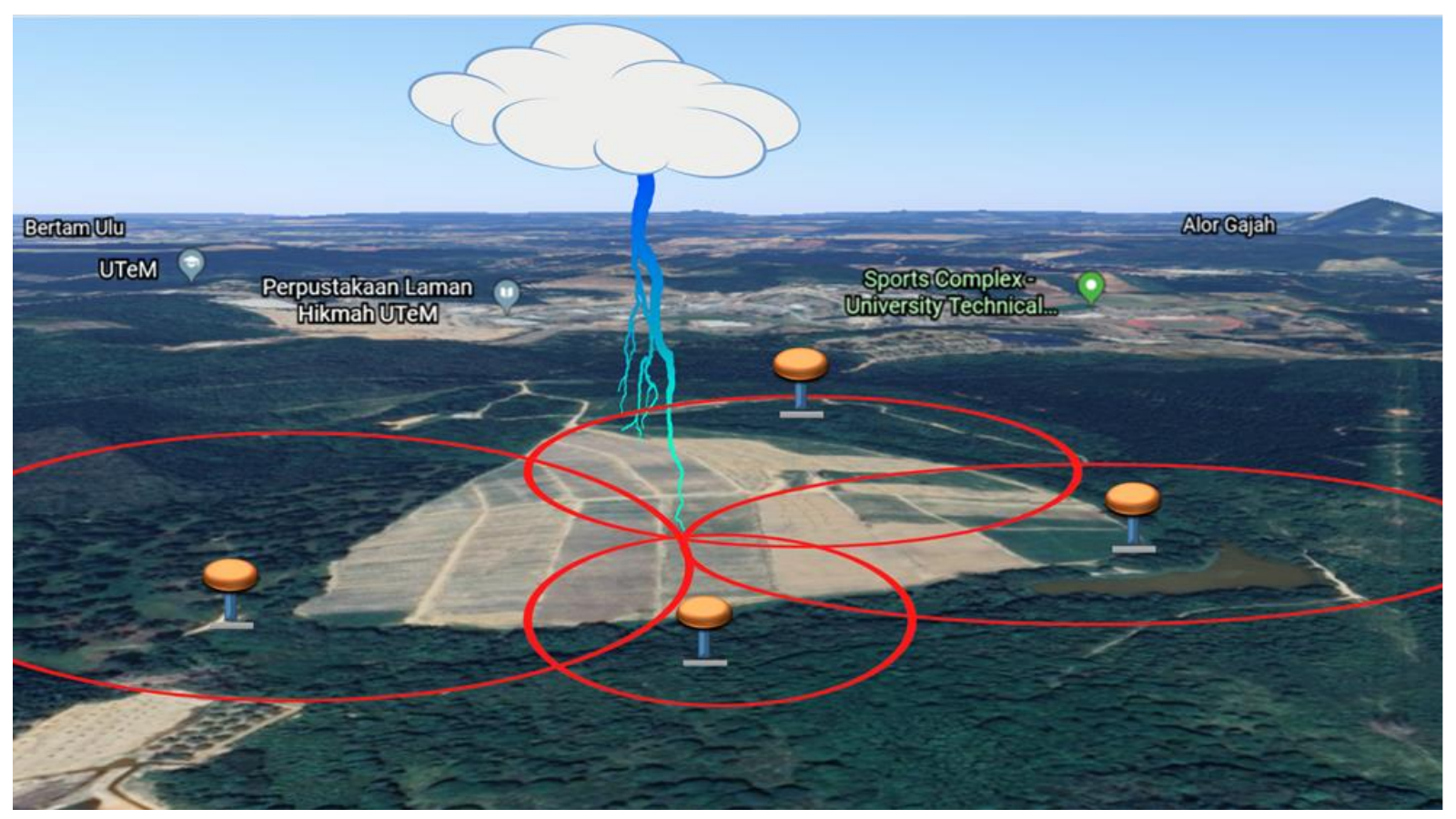}
\caption{Time of Arrival lightning localization technique.}
\label{Fig:Fig4}
\end{figure}

The TOA is generally composed of multi stations that utilize the information obtained from the time-of-arrival data to construct a three-dimensional mapping of lightning flashes. The first study of the TOA lightning technique was developed in Florida by Proctor \cite{Proctor1981}, which works with only single-short baselines and can provide the time for some portions of the lightning electromagnetic (EM) field signal. Later, Lennon and Maier, \cite{Lennon1991} used the same system of \cite{Proctor1981}, and attempted to locate the lightning sources from CG and in-cloud flashes in 3D. Two different systems were operated at a frequency range of 60-300 MHz with an initial operating frequency of 63 MHz for the first system and 255 MHz for the second. Hence, the system was built based on two independent antenna arrays to provide a fast data quality check in real-time. The intention was to design a Lightning Detection and Ranging system (LDAR) for the location mapping of CG and in-cloud flashes of providing a 3D location of RF pulses for each lightning flash. This kind of system could provide more than a thousand 3D locations within each lightning flash. It is similar to the system introduced by Proctor \cite{Proctor1984}, but data acquisition is automatic, and the data displayed are generated in real-time \cite{Rhodes1989}. 

Consequently, Cummins \textit{et al.} \cite{Cummins1998} continued the study of the TOA technique. However, their system was intended to upgrade the design and implementation of the National Detection Lightning Network (NLDN), with data from TOA / MDF studies. The main target was to improve the system location accuracy, system efficiency, and detection, as well as to estimate the peak current of all the strokes in the ground flashes to meet the needs of all-electric utilities. In this system, the delivery of both stroke and flash information in real-time was the main focus. The combination of the two techniques (TOA and MDF) showed a noticeable improvement in lightning detection efficiency and produced better results in terms of reliability compared to NLDN. Hence, both studies preferred to use two techniques for measuring the arrival times and directions of all the strokes. Further studies required changes in the way the data interpreted. Cummins \textit{et al.} \cite{Cummins2000a} came up with a more useful understanding of the combination of the two methods based on typical lightning strokes in Florida, which were detected using five sensors of NLDN. A combination of direction finding and time-of-arrival was the primary method for lightning localization to obtain more information about azimuth, area, altitude, and discharge time. 

Thomas \textit{et al.} \cite{Thomas2004} upgraded the system of TOA measurement of lightning in South Africa as pioneered with measurements based on the 3D mapping. The study focused on the detailed breakdown of individual lightning discharges based on five-station arrays. Perpendicular baselines were implemented for further studies of the individual lightning discharge of the breakdown. Therefore, the centre frequency was set at 60 to 66 MHz. The system was first tested in the New Mexico Tech LMA, with the time of uncertainty empirically found to be about 30 root mean square (RMS), corresponding to source location accuracies as good as 10 m RMS. Due to the different numbers of station arrays, the TOA measurements were not capable of eliminating location uncertainties, and the accuracy of the location was treated in some events outside the system network. Therefore, systematic errors in lightning observations for LMA were found, around 43 ns RMS of the defined pulse shape.

Amir and Ibrahim \cite{Amir2012} introduced a multi-station system using a short baseline for VHF and TOA technique. This study focused on the connectivity of three antennas with a distance of 10 m through a national instrument data acquisition (NI-DAQ) connection to the main personal computer. The main aim was to display the collected data to determine and calculate both azimuth and elevation incident angles. Another target was to determine the exact location of the lightning strokes. To determine the incident angle, a third antenna was added through two baselines perpendicular to each other. Therefore, the captured signal could be monitored through the personal computer by saving the filtered signal to estimate the incident angles using LabVIEW software. A short baseline was applied based on the TOA method to perform the measurements and convert them into 2D. As a result, the CG lightning strike data detected from the plate of the antenna could be analysed and defined. Localization of the multi-station short baseline of the alarm system was done using three broadband antennas. Due to the slow speed of the DAQ, their system can only provide a short recording time; therefore, measurements require a longer time to record the collected data.

 Liu \textit{et al.} \cite{Liu2020} study conducted at the north of Jiangsu Province, China using the TOA technique. The first study proposes a 3D VHF lightning mapping system using a waveform cross-correlation technique. Furthermore, the time difference with a small period of 10 microseconds used for better TOA positioning. Improvement structure was achieved which include K-process and two overlapped K-processes for gaining a high location accuracy of the TOA system with a limited speed of (1.1$\times$10$^7$ m/s). The system VHF bandwidth was 5 MHz with adjustable centre frequency according to the local VHF electromagnetic environment. The system operated based on five substations with a centre frequency of 63 MHz, while the sixth antenna operated at a centre frequency of 53 MHz owing to the local interference. The system was limited to a low sampling rate of 20 MS/s, and additional sensors are required to increase measurement accuracy and waveform TOA method involved to analyse the lightning cases of interest further and still in imperfection stage.
 
Table~\ref{tab2}. summarizes the studies related to TOA lightning mapping. The table depicts the purpose and description of each method. For each study, the contribution and main limitations are highlighted. 

\hphantom{
\cite{Proctor1984}
\cite{Lennon1991}
\cite{Cummins1998}
\cite{Thomas2004}
\cite{Amir2012}
\cite{Edens2012}
\cite{Kawasaki2012}
\cite{Liu2020}
\cite{Quan2003}
\cite{Sun2013}
}
\vspace{-0.18in}
\begin{sidewaystable*}
\caption{Summary of related work of (TOA).}
\begin{tabular}{l}
	\begin{minipage}{1\textwidth}
      \includegraphics[width=\linewidth, keepaspectratio]{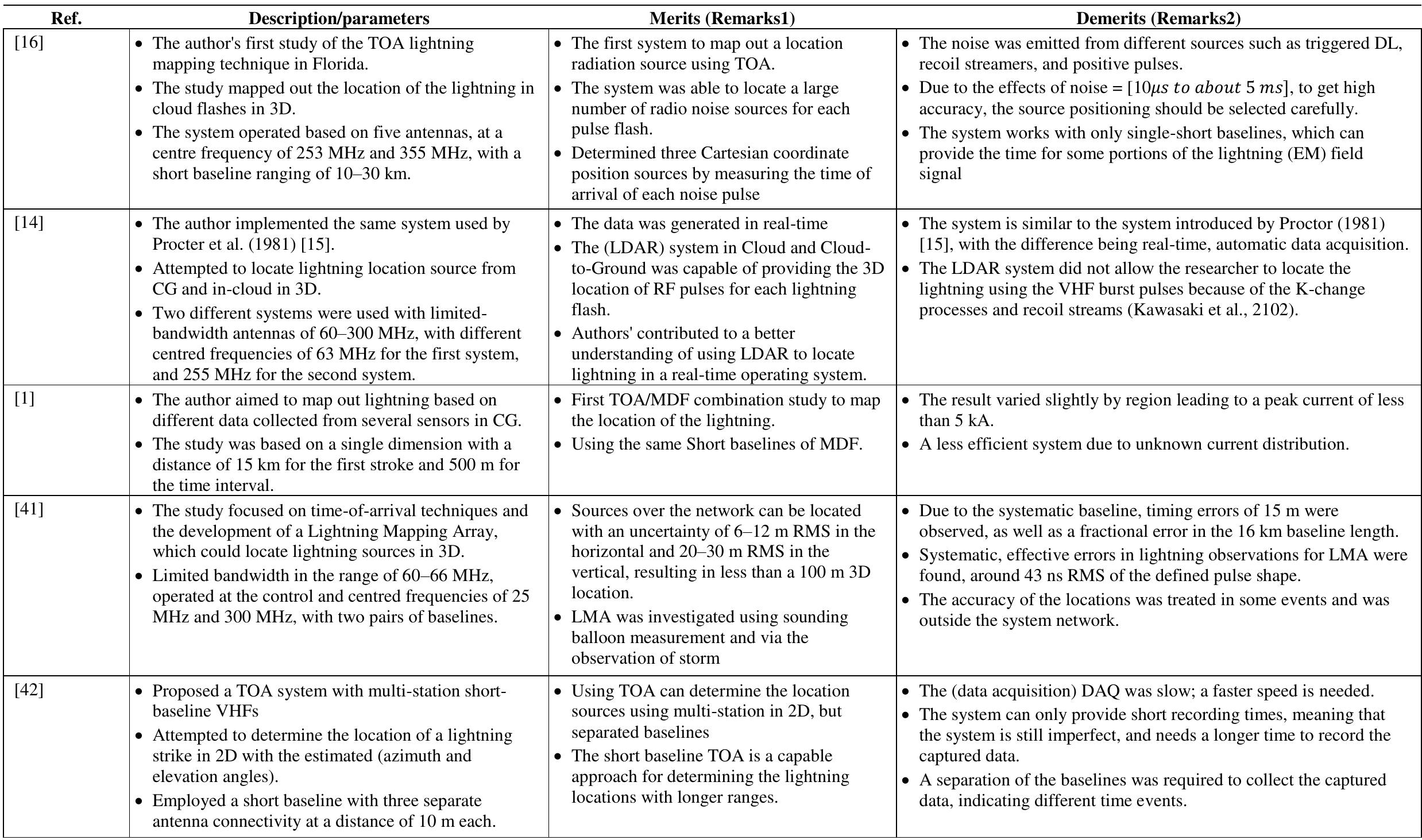}
    \end{minipage}
\end{tabular}
\label{tab2}
\end{sidewaystable*}

\begin{sidewaystable*}
\begin{minipage}{1\textwidth}
	\includegraphics[width=\linewidth, keepaspectratio]{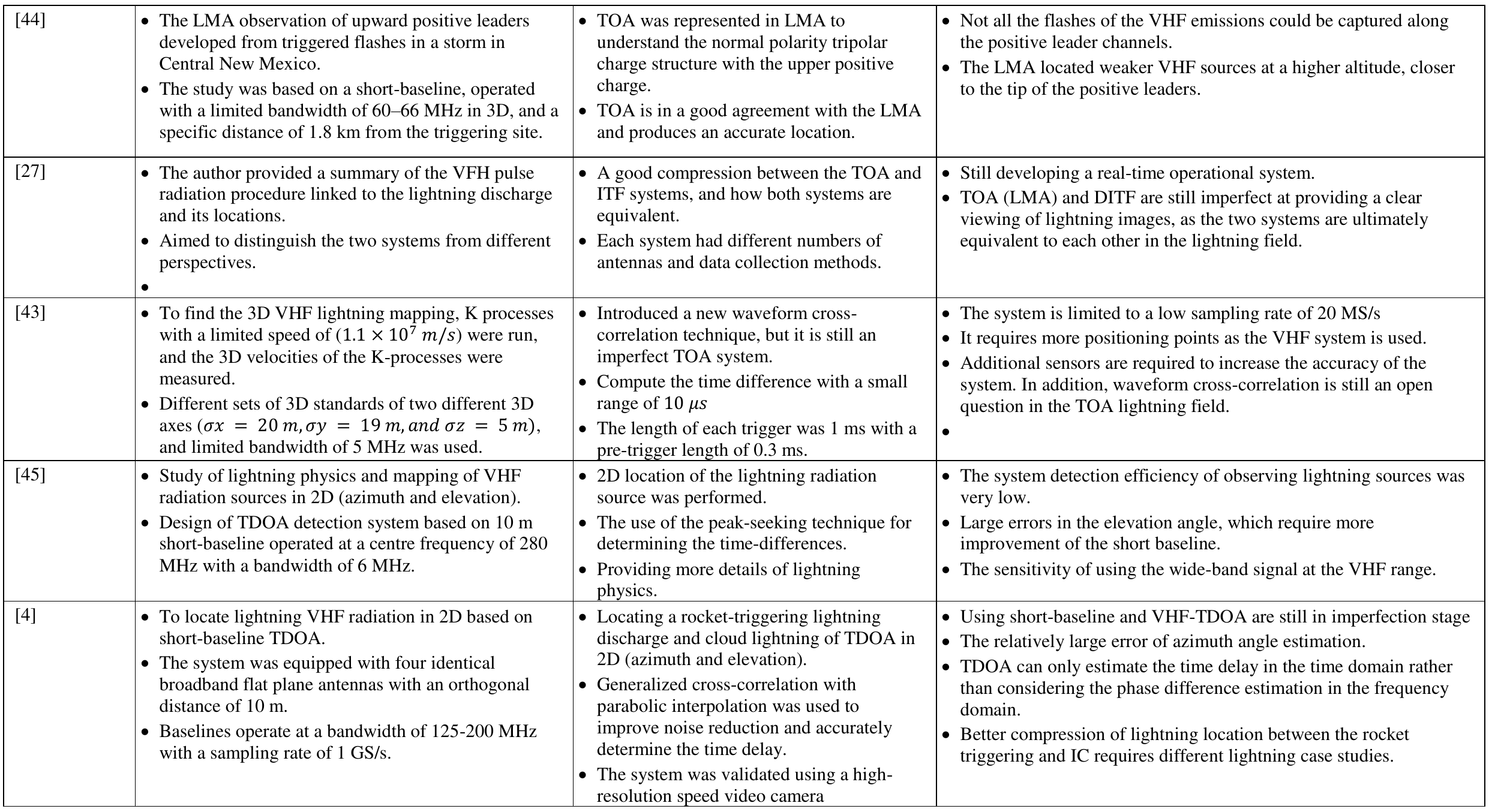}
\end{minipage}
\end{sidewaystable*}

The new measurements based on TDOA, mostly using short baselines, while the real difference depends on the number of antennas. Based on this, Sun \textit{et al.} \cite{Sun2013} introduced a new localization system using four extra fast/slow antennas to measure the electric field changes applied in the global positioning system (GPS) system. TOA systems accurately measure the arrival times of impulsive VHF at numerous ground locations, they are widely spaced over distances as high as tens of kilometres. TOA fails to produce accurate estimations of lightning radiation when short-baselines are used. However, for a combined TOA/TDOA, it improves the configuration of the antenna system allowing the use of short-baslines. Unlike TOA, the TDOA does not require to implicitly record the arrival time of lighting radiation; instead, it uses signal processing methods to estimate the time delay of the received two signals at each baseline (short-baseline) \cite{Quan2003,Sun2013}. 

For the very short-baseline system, the TOA usually operates at a VHF band with a frequency range of 30-300 MHz and with a receiving frequency range of approximately (30-100) MHz \cite{Rakov2013}. The very short-baseline system consists of two or more VHF-TOA receivers with a spacing and time difference between the arrivals of the individual VHF pulses of approximately (1-100) microsecond. As a result, although multiple antennas are used to detect the VHF pulses, the pulses could become challenging to distinguish. To overcome this issue, the very short-baseline technique, which uses closely spaced antennas with almost identical positions and with all the pulses received by antennas, can be applied. Meanwhile, the short baseline of the TOA for a three-dimensional location requires at least four antennas to obtain a very high imaging accuracy for the lightning sources. Overall, the very short-baseline technique is useful for estimating the azimuth and the elevation of VHF sources. In contrast, the short-baseline technique is useful for developing two independent systems and providing EM images to develop the channels for any type of lightning flash \cite{Kawasaki2012, Oetzel1969}. Besides, Lewis stated that by using the long-baseline, the TOA system could operate at a VLF/LF with a bandwidth of (4-45) kHz, separated by over 100 km and spread over four stations. The problem with the long-baseline system, however, is the difficulty in identifying the same pulse features in the signal received by different antennas.

\section{INTERFEROMETER}
The basic principle of the ITF system is to estimate the phase difference of the electromagnetic VHF emissions detected by closely spaced pairs of antennas. The TDOA or phase difference of incident electromagnetic pulses is estimated for every two antennas (baseline) separated by distance $d$. As shown in Figure 5, the pair of antennas BC and BD are used to calculate the phase difference between the implemented antennas where antenna B is the reference. Then the azimuth and the elevation angles of the radiation sources can be derived from the ITF geometry.

\begin{figure}[!h]
	\centering
	\includegraphics[width=0.4\linewidth,keepaspectratio]{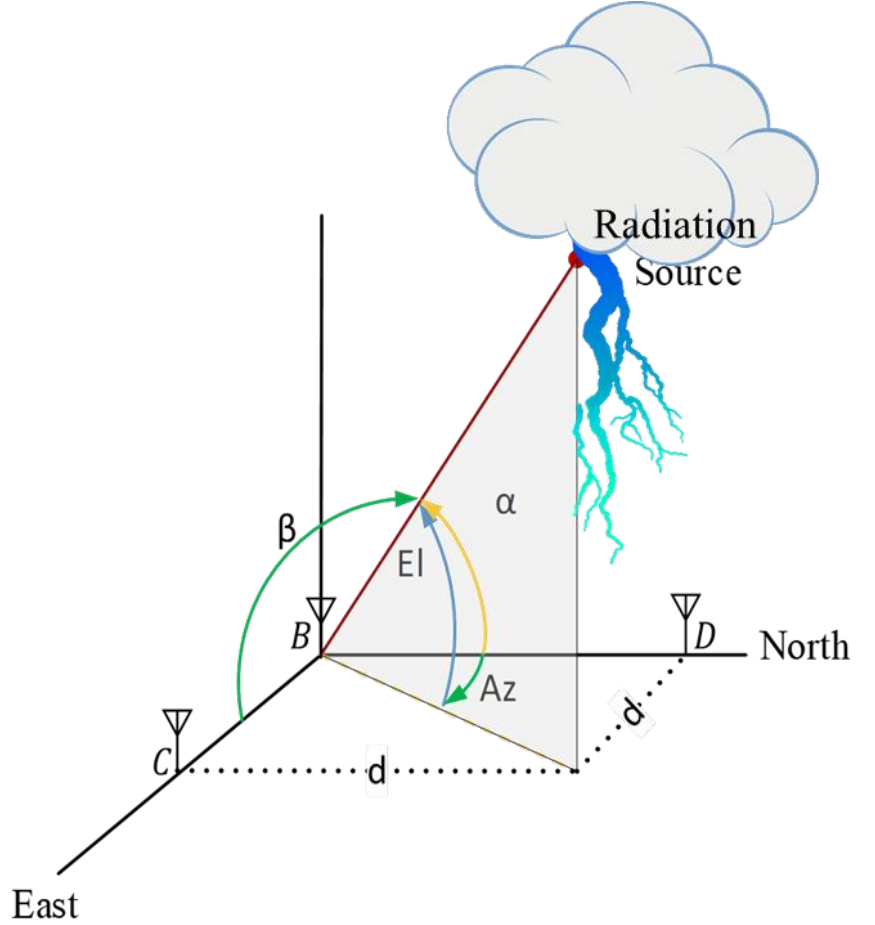}
\caption{Interferometer system.}
\label{Fig:Fig5}
\end{figure}

The existing lightning ITF systems can be categorized into two types according to their operating frequency, the narrowband ITF, and broadband ITF. The broadband ITF is superior to the narrowband in which it can record lightning emissions in a higher resolution providing more information on temporal lightning events. A single broadband ITF station can provide a continuous lightning mapping in two-dimensional where a synchronous multi-station system can visualize the lightning progression in three-dimensional maps \cite{Liu2018}. The broadband lightning ITF method was developed rapidly in the field of localization and is considered a novel tool in lightning discharge investigations \cite{Mardiana2006, Shao1996, Ushio1997}. The basic idea of the broadband ITF is to estimate the phase difference at various frequency components of Fourier spectra between a pair of broadband antennas. 

Warwick \textit{et al.} \cite{Warwick1979} were the very first to implement a lightning ITF based on a single baseline operating over a narrow band with a centre frequency of 34 MHz with two VHF antennas separated by 80 m. The authors found that the instrument was useful for tracking the motion of lightning by using Quadrature Phase detection method to determine the phase difference of arrival. However, working with only a single baseline ITF, the system was not able to produce maps or images of the lightning source, and some fringes with electric field recorded by lightning flashes. Therefore, the ITF system still in the imperfection stage, which still requires further improvements for locating a more extended portion of the radiation emitted by lightning. Later, Hayenga and Warwick \cite{Hayenga1981} conducted another study as a first system to estimate the lightning mapping in 2D (azimuth and elevation) location based on ITF with 2.5 $\mu$s resolution. The system operates at 34.3 MHz narrow-band (3.4 MHz) with two pairs of antennas separated by 15 m. Instruments were implemented perpendicular to crossed baseline instead of a single baseline. This is where the azimuth and elevation incidence angles were determined. However, due to the amount of errors in the phase difference measurements, the authors employed a baseline length of 2$\lambda$ for better localization. As a result, the findings showed a breakdown progression during lightning flashes, but this did not improve the ITF technique. Nevertheless, the study was the first to apply a 2D ITF. Therefore, the system was susceptible to a phase ambiguity, fuzzy maps of lightning flashes, multiple baselines are required to eliminate the ambiguities and provide more source information (e.g. \cite{Morimoto2005, Yoshida2012}). 

In general, the methods for mapping lightning using broadband digital ITF can be classified into three types according to the processing technique, linear fit method (phase fitting), wavelet transform-based techniques, and cross-correlation-based techniques. Rhodes and Krehbiel \cite{Rhodes1989} continued the study of ITF lightning providing more advanced with an increased operating frequency from 34 MHz to 274 MHZ. Two pairs of baselines, short baseline of 1/2$\lambda$ and long-baseline of 4$\lambda$ instead of a single pair of crossed baselines 2$\lambda$, to improve the angular uncertainty and to avoid the fringes ambiguity. The study focused on CG lightning using analogy narrow band mixers. The results were measurements based on a 2D (azimuth and elevation) with 1 $\mu$s time resolution direction of arrival in the radiation source. The long-baseline produced noticeable high accuracy results, whereas the short baseline was not useful as it determines some portion of the flashes compared to the long-baseline. 

Shao and Krehbiel \cite{Shao1996} show study improved version of the ITF system used by \cite{Rhodes1994}. The antenna configurations of short and long baselines were changed to 1$\lambda$ and 4.5$\lambda$ and limited broadband of 40-350 MHz. Furthermore, integer arithmetic to convert phase measurements to azimuth and elevation values were attempted. Due to the noise immunity of the system, combining the baselines by an automated process, which provides a comparative analysis of both CG and IC flashes of lightning with increasing spacing between antennas helped to resolve the fringe ambiguity. The real-time lightning locations could also be displayed by focusing on broadband ITF. Throughout 1995 and 1996 the authors concentrated on the development of the ITF with a full description of CG and inter/intra-cloud (IC) flashes based on the advanced of an affordable broadband ITF. However, the study identified a number but not all of the features of IC discharges, also there still some small errors in the short baseline measurements. 

Mazur \textit{et al.} \cite{Mazur1997} came up with two different systems; lightning detection and ranging interferometric ``LADR'' (TOA) and French office national d'etudes et de Recherches Aerospatiale  ``ONERA-3D'' (ITF) to locate CG lightning source at 2D (azimuth and elevation) maps. Both systems operated at two different centre frequencies LDAR (66 MHz), and ONERA-3D (110-118 MHz). LDAR data bandwidth was 6 MHz, where only 1 MHz for ONERA-3D. The study focused on the comparative of simultaneous lightning mapping of the same storm using LDAR and the ONERA-3D systems. IC and CG lightning events were analysed. ONERA-3D maps best the fast intermittent processes and dart leaders while LDAR maps best the continuously developing intra-cloud processes. Therefore, the LDAR (TOA) system failed to detect positive leaders and slow negative breakdown, at the same time LDAR does not perform well in mapping radiation sources associated with stepped and dart leaders propagating toward the ground. LDAR is similarly inadequate to map K changes inside the cloud, while the ONERA (ITF) system does not map radiation from continuous positive leaders.

Shao and Krehbiel \cite{Shao1996} introduced a new technique called linear fit, which focuses on the change of phase difference with frequency. It based on a single vertical baseline to obtain further detailed measurements of the descent of a leader towards the ground. Data were recorded continuously for short durations just before the return strokes. Based on these procedures, the manufactured ITF was able to locate and reconstruct lightning progression in a one-spatial dimension. In addition, as a proof concept, the study presented a new method to determine the time difference between antennas.

Ushio \textit{et al.} \cite{Ushio1997} used the same linear fit technique, extending the antenna system with a pair of orthogonal horizontal baselines. As a result, two baselines allowed the lightning to be mapped in 2D (azimuth and elevation). Due to the higher digitization rate, the broadband ITF faced difficulties in recording the emitted radiation from the lightning discharge. Therefore, the authors developed triggering techniques to better record the radiation. The VHF broadband digital ITF was developed with measurements to correlate with both the amplitude and phase of the receiving signal via pairs of antennas. 

The earliest instruments for reconstructing CG lightning discharge were developed in Australia in December 1997 to address the most significant lightning spectrum bandwidth and the limitation of digitized memory storage. More improvements were made by \cite{Mardiana2000} who used triggering techniques for recording EM pulse data with a frequency between 0-250 MHz. The focus was on the location of fast-EM radiations emitted from lightning discharges, and on acquiring data on the emitted flashes of lightning. Therefore, the ITF technique was able to estimate and extract the phase difference at variable frequencies using a discrete Fourier transform (DFT). The signal was delivered in the frequency domain of the incidence of EM pulses at two antennas. As a result, the azimuth and elevation angles were resolved, and strong EM pulses were detected along with the lead progression, where the best reconstruction of CG lightning discharge was displayed in two dimensions and in time sequences.

Mardiana and Kawasaki \cite{Mardiana2000} attempted to solve the issues of large lightning frequency bandwidth using a high digitization rate and equational triggering techniques due to the limitation of digitized memory capacity. Several attempts were made to solve this problem by increasing the frequency range from 25 to 250 MHz with the same perpendicular baselines and the same mathematical techniques. This study added another antenna to calculate the azimuth and elevation incidence angles using cross-correlation and Fourier transform techniques between two antennas. Based on the changes made, the results showed notable success in recording data for broadband fast-moving EM as well as in reconstructing CG lightning discharges. Moreover, it was observed that broadband ITF was more useful than the narrowband for locating heavy lightning.

Dong \textit{et al.} \cite{Dong2001} used a new comparison to investigate the emitted radiation and to study the development of the decisive leader in artificially triggering lightning by using a power spectrum with a frequency of 25 MHz of a high-pass filter. Different frequencies of the positive leader with a range of 25 to 30 MHz and negative breakdown processes of 60 to 70 MHz were tested. Furthermore, both narrowband and broadband ITF were implemented to identify which was more useful and stronger for emitted radiation detection. As a result, narrowband ITF from the IC produced stronger radiation. In addition, due to the intermittent sampling of the broadband, ITF was not able to record the entire radiation by triggering lightning. The authors claimed that using a high pass filter close to lightning produced radiation from the positive leader rather than broadband ITF. Therefore, recording at a high sampling rate was required for better resolution of radiation signals from the antennas.

Kawasaki \textit{et al.} \cite{Kawasaki2004} conducted a study to improve the broadband ITF of CG, giving more attention to changing the bandwidth range (10 to 250 MHz) by implementing two orthogonal baselines with a specific length of 10 m. The study was based on two pairs of antennas with two independent baselines and amplifiers equipped with antennas digitized at a sampling rate of 500 MHz and 8-bit resolution. To obtain 3D lightning, they developed a broadband ITF for imaging the lightning channel and identified the positive charge distribution inside a thundercloud. As a result, different charges of the thunderclouds were observed, but not fully identified due to insufficient data for preparing advanced broadband ITF with higher spatial and temporal resolutions.

Mardiana \textit{et al.} \cite{Mardiana2006} used the same baselines as \cite{Kawasaki2004} for data digitization at 500 MHz with the same baseline length of 10 m. The focus of their study was on IC temporal development observed by the VHF broadband of ITF. The IC flashes were characterized by the active and final stages indicated by streamer breakdown at the lower-level channel. The main idea was to observe the most robust radiation sources in the lower part of the storm towards the positive charge region. Overall, a 2D method was applied for observing the IC lightning discharge via the VHF broadband system. A few of bi-directional IC discharges are also represented, but not all the IC lightning discharge features were reported in this study.

Qiu \textit{et al.} \cite{Qiu2009} carried out a study and algorithm using new advanced techniques. The authors focused on reducing the phase noise of VHF radiations. Noise effectiveness was located and found during system implementation. Also, due to the phase difference spectra, some of the random noise was added to the system with noticeable effects on accuracy as well as reliability. Therefore, several algorithms were applied, such as the phase-filtering algorithms, which represented circular correlation and translation-invariants. Based on this method, the problem of simulated and experimental data distorted by extra noise addressed, and the incident angles were resolved. The wavelet transform based on the Myert wavelet function was also used to remove or reduce noise. The experimental results show that better accuracy and reliability were achieved for phase noise reduction, and better location accuracy obtained for incident angle resolution, compared to the conventional approaches. However, the use of wavelet transforms is still an open question in the localization field.

Cao \textit{et al.,} \cite{Cao2010} conducted a study of an ITF system operates at 125-200 MHz consists of four perpendicular broadband flat antennas separated horizontally by 10 m. Fast and slow antennas were used to capture the lightning electric fields. The System was equipped with two GPS receivers to provide accurate trigger time. Differences in arrival times and phase differences were calculated using the geometric model and FFT approach, respectively. The system has resulted in 2D lightning discharge mapping of multiple stroke CG flashes, and Fourier filtering and \textit{Symlet} wavelet denoising were applied to suppress the unwanted high frequencies (>200 MHz). Data was stored using a Lecory oscilloscope with a 1 GHz sampling rate. The sequential triggering technique was used to reduce the data samples. Overall, System used what is referred to as the short-baseline TOA (improved version of TOA) with ITF antennas, noise reduction to improve the lightning source detection and limited measurements to the single of CG flashes. The system did not provide a clear direction for the lightning flashes.

The cross-correlation technique is a well-established technique, although, in the field of lightning mapping, it was first used by \cite{Akita2010, Akita2014}. The authors contributed to the VHF broadband digital ITF system to locate the impulsive VHF radiation sources of lightning discharges \cite{Akita2010}. Akita \textit{et al.,} \cite{Akita2014} introduced further upgrades to the digital ITF system with a new processing method of phase differences distribution analysis to better locate the lightning-related radiation sources. The measurements used two ITF systems to calculate the location of lightning flashes in 3D. Hence, azimuth and elevation angles of the radiation sources were derived. Moreover, measurements concentrated on recording the ground to estimate the charge distribution occurring inside the cloud. A two-ITF mapping system was deployed, approximately 15 km apart. The system reconstructed lightning charges by recording the 3D location of sources, as well as proving its ability to measure the 3D propagation speed of fast-moving ionization waves, termed the K-Change.

The best ITF measurement studies were represented by implementing a change in the frequency bandpass filters from 20-125 MHz to 20-80 MHz, installed inside the antennas to avoid contaminated aliasing \cite{Akita2014}. The authors developed a new phase fitting technique to continuously record the radiation observations using digital interferometer (DITF) and to use processing algorithms for enhanced lightning detection. Meanwhile, Stock \textit{et al.} \cite{Stock2014a} developed another common technique called cross-correlation. At that time, neither study yielded any useful findings of scientific interest, so the authors continued their attempts with different measurements and collection of data to obtain new results. Stock and Krehbiel \cite{Stock2014} redesigned the antenna system, moving it further from the noise generation with an extended baseline at a specific distance of 10 to 16 m; each antenna was reconnected to a single cable. A fourth channel was added to better record the VHF ITF during the use of multiple baselines. In this case, the number of flashes recorded was increased from three during the 2012 storms to over a thousand in 2013, producing better quality results and improving the triggering mechanism for detecting lightning discharges.

Akita \textit{et al.} \cite{Akita2010} used the VHF digital ITF system \cite{Ushio1997} to investigate the progression of K-process (K-type breakdowns) of cloud flashes. Three-dimensional localization of radiation sources was achieved by using two statins of the conventional two-dimensional ITF and taking into account the arrival times. The entire system setup involves two ITF systems separated by 5.2 km implemented in two different sites with one ITF that includes slow E-field antenna. The analysis of two cloud flashes using phase difference estimation reveals more details of K process visualization and progression. However, lightning map validation was not possible due to the lack of real images of in-cloud channels. For that reason, most of the studies implemented cross-correlation techniques rather than phase fitting techniques for lightning mapping. 

Nakamura \textit{et al.} \cite{Nakamura2014} ITF system was built based on a VHF broadband (25-80 MHz) of three antennas, a sampling rate of 200 MS/s., and one additional antenna for triggering signals. A low-frequency system of single fast antenna called broadband observational network for lightning and thunderstorms (BOLT) (500 Hz-500 kHz) was used for lightning localization in 3D. The comparative analysis of both ITF and BOLT was in lightning detection. Therefore, the ITF system was able to detect the small lightning EM waves. Meanwhile, the BOLT system requires very long baselines, which fails to detect huge EM waves in a short time. Hence, the system requires a huge memory to store data.

Stock \textit{et al.} \cite{Stock2014} proposed a hybrid approach where two systems were used, a broadband (20-80 MHz) ITF and TOA Langmuir LMA. The ITF consists of three flat plate antennas separated by 10.2 m, and another fourth antenna for triggering recording. ITF continuous data were recorded with a sampling rate of 180 MS/s to observe the entire lightning activity. A generalized cross-correlation in the frequency domain with windowing functions was used to determine the segmental TDOA between the implemented antennas. The authors combined 2D-ITF maps with 3D-LMA to provide a quasi-3D lightning map. The use of LMA with ITF provides a superior 3D spatial resolution instead of using multiple stations. Overall, both systems ITF and LMA are sensitive to environmental noise levels and require post-processing to eliminate the noise and non-valid localizations. Therefore, the use of windowing (1.52 $\mu$s) with an overlap of 98\% to obtain the time delays shows an estimated angular uncertainty of 0.82$\degree$  for azimuth and 0.63$\degree$ for elevation angles.

Abeywardhana and Fernando \cite{Abeywardhana2018} aimed to locate the VHF radiation source in 2D (azimuth and elevation). A VHF broadband (10-80 MHz) ITF system with three flat circular capacitive antennas arranged orthogonally and spaced of 10 m, centred frequency of 45.5 MHz. Tektronix MDO-3034 oscilloscope used to digitize and store the data with a sampling rate of 250 MS/s. The fourth antenna was added to record the fast-electric field. The cross-correlation method was used for mapping the lightning progression. The system resulted in a 2D lightning source, and maps validated with the visible events of CG flashes using a high-resolution camera. Data recording length was limited to 40 ms only due to the high storage requirements. Therefore, the ITF system still required some improvements to eliminate the noise with minimum information loss to enhance the lightning maps.

Some of the advantages of using the ITF system toward the TOA are: (1.) No identification of the individual pulses, since the ITF measures the phase difference between narrowband signals corresponding to these noise-like bursts received by two or more closely spaced sensors \cite{Rakov2013}; (2.) using less number of antennas in the closely spaced distance to estimate the TDOA or phase differences between the correlated signals arriving at two antennas. The use of the advanced broadband ITF technology, ``this trend is made possible by the advantage of affordable broadband RF and digital signal processing electronics'', as the ITF required to be integrated over windowing \cite{Mardiana2000, Shao1996, Stock2014}; (3.) Generally, the ITF is a system used for determining the direction of the radiation source form a combination of phase measurements in a relatively small or wide bandwidth. The antenna system is made up of an array of dipoles in which are connected to an ITF receiver to determine the phase difference of the received observations. The phase difference usually obtained by comparing the received signals from several antennas or by comparing these signals with a standard local oscillator. The accuracy of phase measurements is then achieved by integrating the desired time resolution. Finally, by post-processing, the phase differences, the 2D or 3D network (azimuth and elevation) is obtained for the radiation source. The radiation triangulation mechanism is calculated in the central processor for radiation events localization using the received information from several sensors \cite{Lojou2009}.

Table~\ref{tab3} summarizes studies of the ITF, showing how the ITF rapidly developed based on different lightning measurements. The main difference of the previously proposed ITF systems depends on the range of bandwidth used, the type of baseline (vertical, orthogonal, or short), and the wavelength of each measurement (the most changeable parameter), the number of antennas, and the distance between the implemented and designed antenna. Some of the similarities cluster around using the ITF technique for lightning localization, which can locate VHF sources more successfully than previous non-ITF methods. The dimension in which the ITF was conducted along with the number of antennas used in each mapping system is indicated. For each study, Table~\ref{tab3} includes remarks to reflect the newly implemented methods and system advantages. Other remarks of limitations and disadvantages are also illustrated. Also, most of the authors proposed different lightning mapping methods and various solutions to gain better lightning localization based on different setups with different data collected. Finally, the broadband ITF currently in use presents a significant improvement over other lightning ITF, compared to the TOA.

\hphantom{
\cite{Warwick1979}
\cite{Hayenga1981}
\cite{Rhodes1989}
\cite{Rhodes1994}
\cite{Mazur1995}
\cite{Shao1995}
\cite{Shao1996}
\cite{Mazur1997}
\cite{Ushio1997}
\cite{Cao2010}
\cite{Akita2010}
\cite{Sun2013}
\cite{Stock2014}
\cite{Nakamura2014}
\cite{Abeywardhana2018}
}
\vspace{-0.18in}
\begin{sidewaystable*}
\caption{Summary of related work of (ITF).}
\begin{tabular}{l}
	\begin{minipage}{1\textwidth}
      \includegraphics[width=\linewidth, keepaspectratio]{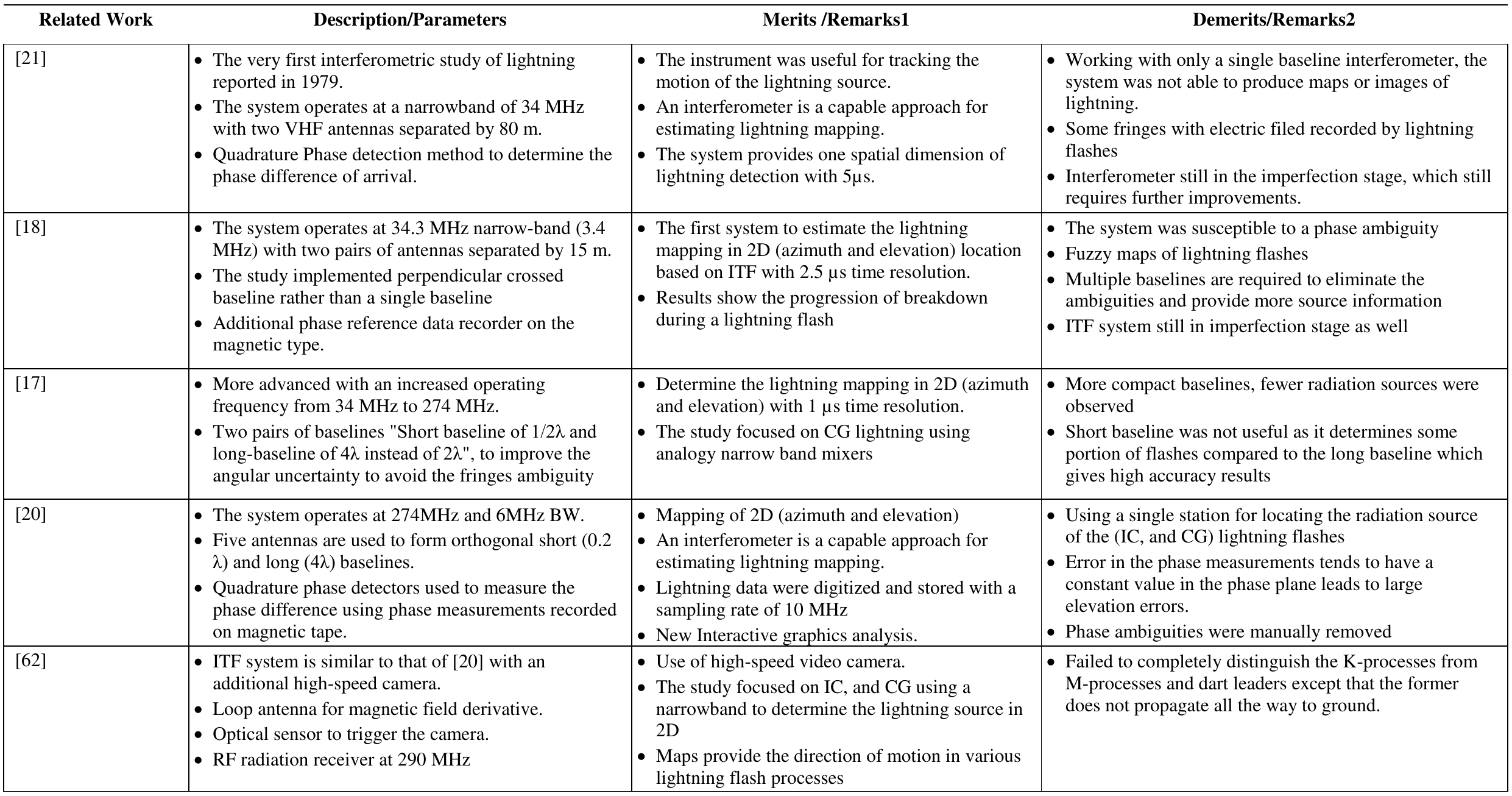}
    \end{minipage}
\end{tabular}
\label{tab3}
\end{sidewaystable*}

\begin{sidewaystable*}
\begin{minipage}{1\textwidth}
	\includegraphics[width=\linewidth, keepaspectratio]{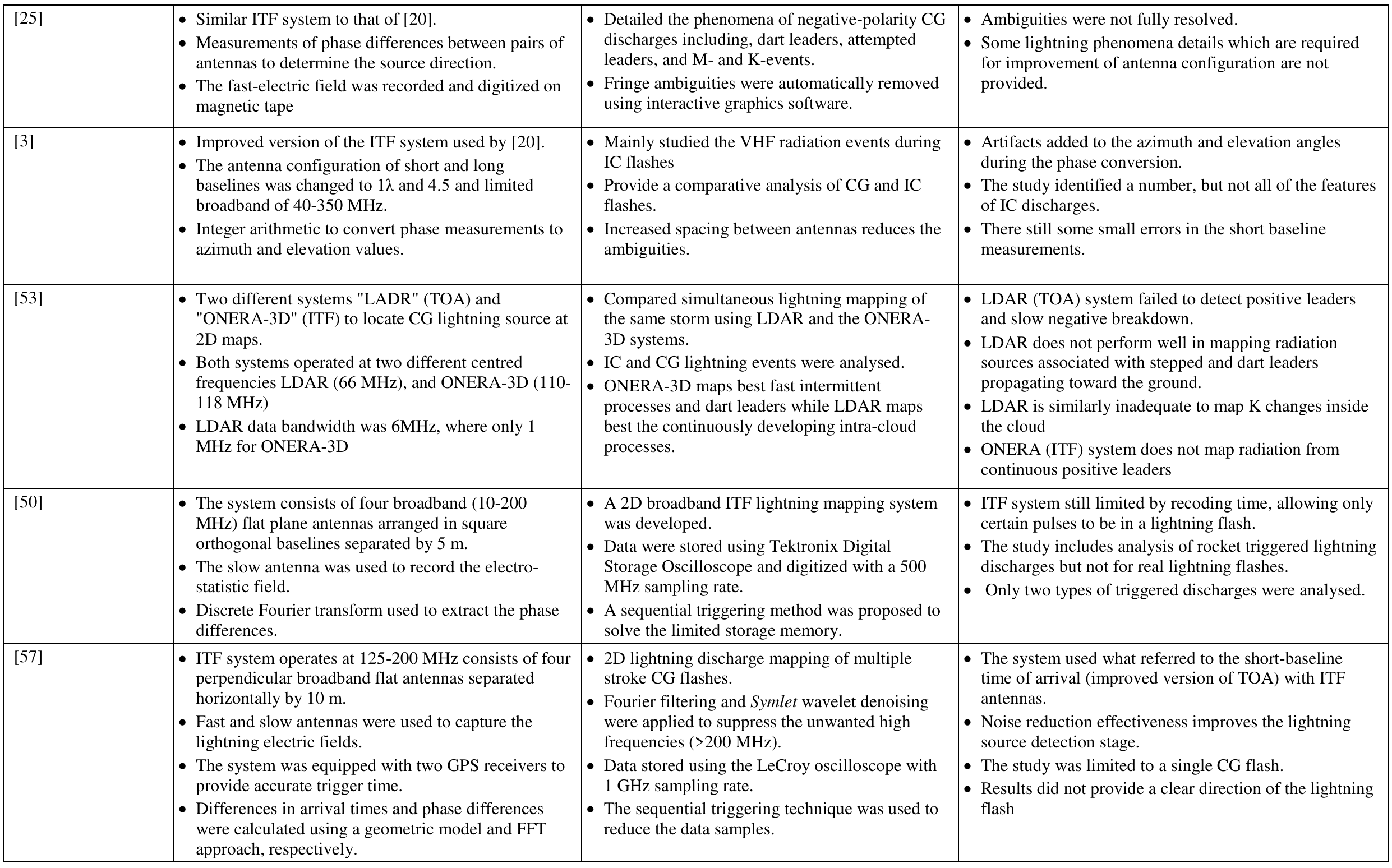}
\end{minipage}
\end{sidewaystable*}

\begin{sidewaystable*}
\begin{minipage}{1\textwidth}
	\includegraphics[width=\linewidth, keepaspectratio]{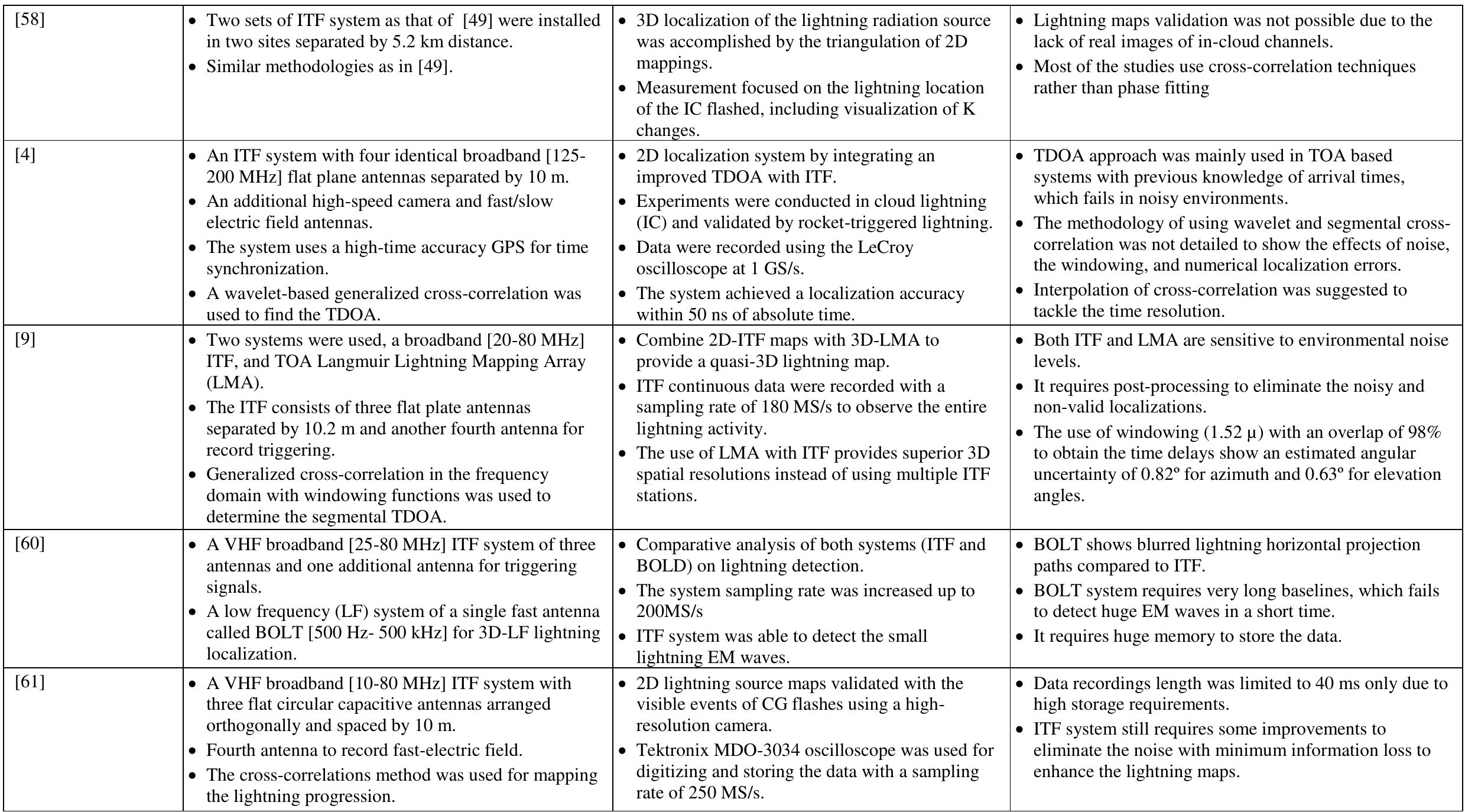}
\end{minipage}
\end{sidewaystable*}

\section{DISCUSSION AND OPEN ISSUES}
Technically, a major challenge in using conventional methods to map the lightning sources is that they are highly dependent on the pre-processing steps. There is still room to improve the performance of estimating lightning mapping of ITF systems if better pre-processing techniques are utilized. One key element of the pre-processing improvements is the use of more reliable methods to remove the unwanted noise and give a precise estimation of the time difference of arrival (TDOA). There are a couple of ways in which this can be performed. First, the lightning signal can be filtered and prepared for further signal processing. It requires the use of various filters to completely remove the noise that is superimposed on the lightning signal during the data collection and can induce a TDOA estimation error, which will affect the mapping performance. Therefore, a possible research direction that we intend to follow is to investigate and benchmark the use of various filtration techniques by introducing wavelet transform (WT), Kalman filter (KF), and bandpass filter (BPF), where each type of filtration has its mathematical representation. 

Secondly, different cross-correlation methods can be compared in three domains; time-domain (TD), frequency domain (FD), and wavelet domain (WD). Each method of cross-correlation is implemented using interpolation and resampling techniques. Previous studies achieved a remarkable accuracy by using cross-correlation for better lightning mapping. According to Stock \textit{et al.,} \cite{Stock2014} cross-correlation is the primary technique to map the VHF lightning signal, although it has a relatively poor performance of 60\% accuracy to estimate the lightning mapping. This opens the door for further improvement, by introducing new cross-correlation methods (e.g. wavelet-based cross-correlation) in the signal post-processing phase. The lightning mapping estimations obtained using these cross-correlation techniques, may not mimic the natural behavior of real lightning or achieve an optimal estimation of the lightning maps. Furthermore, lightning mapping has many graphical shapes and representations, and it is difficult to judge the accuracy of the representation of real lightning unless there is a source of validation. To overcome this issue, a high-speed camera can be used to validate the lightning results that are presented in elevation and azimuth x-y axis coordination.
 
Finally, observations of broadband VHF pulses associated with lightning discharges by multiple antennas and/or sites are applicable for both TOA and ITF to locate the source positions. To estimate the TOA, the cross-correlation method is occasionally adopted, and there are significant contributions to interpret the mechanism of lightning progression. However, the shape of recorded broadband pulses consists of several peaks that look like a kind of oscillation, and some of the peaks show nearly the same amplitude. Then if we apply the cross-correlation simply, the estimated time difference may not always be correct, though mainly estimation is going well. From the aspect of science, these estimation errors may create some confusion in the lightning physics field such as the imaging of the progression of positive breakdowns.

\section{Conclusion}
Lightning mapping is crucial in lightning monitoring, tracking, detection, and mapping of both CG and IC flashes. This review has presented an intensive discussion of the most related work to lightning detection methods. We discussed the common methods used for lightning mapping namely, MDF, TOA, and ITF. We conclude that the following points are essential for lightning mapping systems: Measurement improvements of radiation sources to produce more accurate mapping, especially for the weak sources; increasing the accuracy of the measurements of both current amplitude and lightning mapping efficiencies, and distinguishing the type of flashes. From this review, we also conclude that VHF is the most common frequency band used for lightning mapping which provides more information on broader frequency components and allows for more accurate mapping in comparison to other frequency bands.

From the discussed literature research work, it is noticed that some of the methods perform better when combined with another one as a hybrid system. This is very clear when a combination of MDF and TOA methods was used which shows significant improvement when compared to using either method alone. We also can conclude that achieving a very accurate mapping using the TOA method requires installing a large number of sensors which can be costly. In contrast, ITF could accurately map the lightning signals with three to four antennas only. The ITF is one of the most promising techniques in mapping signals and we note that with better signal processing, the ITF can achieve high accuracy in mapping the lightning discharges. One good advantage of ITF is that it is not sensitive to noise that is captured along with the recorded lightning signal.

%


\bibliographystyle{IEEEbib}
\bibliography{references}

\end{document}